\makeatletter
\declare@file@substitution{revtex4-1.cls}{revtex4-2.cls}
\makeatother

\documentclass[twocolumn]{aastex631}
\usepackage{mathtools}

\shorttitle{An efficient tidal dissipation mechanism via stellar magnetic fields}
\shortauthors{Duguid et al.}

\graphicspath{{./}{figures/}}

\begin{document}

\title{An efficient tidal dissipation mechanism via stellar magnetic fields}

\correspondingauthor{Craig D. Duguid}
\email{craig.d.duguid@durham.ac.uk}

\author[0000-0003-1199-3469]{Craig D. Duguid}
\affiliation{Department of Mathematical Sciences, 
Durham University,
Upper Mountjoy Campus,
Durham,
DH1 3LE, UK}

\author[0000-0002-6272-9839]{Nils B. de Vries}
\affiliation{School of Mathematics, University of Leeds, Leeds LS2 9JT, UK}

\author[0000-0002-7635-9728]{Daniel Lecoanet}
\affiliation{
CIERA, Northwestern University, Evanston, IL 60201, USA}

\author[0000-0003-4397-7332]{Adrian J. Barker}
\affiliation{School of Mathematics, University of Leeds, Leeds LS2 9JT, UK}

\begin{abstract}
Recent work suggests that inwardly propagating internal gravity waves (IGWs) within a star can be fully converted to outward magnetic waves (MWs) if they encounter a sufficiently strong magnetic field.
The resulting magnetic waves dissipate as they propagate outward to regions with lower Alfv\'{e}n velocity.
While tidal forcing is known to excite IGWs, this conversion and subsequent damping of magnetic waves has not been explored as a tidal dissipation mechanism.
In particular, stars with sufficiently strong magnetic fields could fully dissipate tidally excited waves, yielding the same tidal evolution as the previously-studied ``travelling wave regime''.
Here, we evaluate the viability of this mechanism using stellar models of stars with convective cores (F-type stars in the mass range of $1.2$-$1.6M_\odot$) which were previously thought to be weakly tidally dissipative (due to the absence of nonlinear gravity wave breaking). 
The criterion for wave conversion to operate is evaluated for each stellar mass using the properties of each star's interior along with estimates of the magnetic field produced by a convective core dynamo under the assumption of equipartition between kinetic (convective) and magnetic energies. 
Our main result is that this previously unexplored source of efficient tidal dissipation can operate in stars within this mass range for significant fractions of their lifetimes.
This tidal dissipation mechanism appears to be consistent with the observed inspiral of WASP-12b, and more generally could play an important role in the orbital evolution of hot Jupiters -- and to lower mass ultra-short period planets -- orbiting F-type stars.
\end{abstract}


\section{Introduction} \label{sec_intro}

Tides are important for driving the spin and orbital evolution of short period planets and binary stars. In particular, the rates of evolution due to tidal interactions are governed by how efficient various physical processes are at dissipating the tidal flows.
Tidal dissipation is typically parametrised by the (modified) tidal quality factor $Q^\prime$, \citep{ogilvie_tidal_2014} where smaller values mean more efficient dissipation and hence more rapid tidal evolution. 
For stars with radiative cores, an important source of tidal dissipation comes from the nonlinear breaking of tidally excited internal gravity waves \citep{goodman_dynamical_1998,ogilvie_tidal_2007,barker_internal_2010,barker_three-dimensional_2011}. Excited at the convective-radiative boundary of a convective envelope due to the tidal forcing, internal gravity waves (IGWs) propagate radially inwards through the radiative zone towards the 
centre of the star.
These IGWs are fully damped if their amplitude is large enough to overturn the stratification within the radiative zone, which typically occurs due to geometrical focusing close to the centre.
Upon breaking, the deposition of energy spins up the star from the inside out, forming a critical layer which expands as subsequent IGWs are absorbed.
For stars that 
possess a convective core instead (in the mass range ${\sim}1.2$-$1.6 M_\odot$), IGWs due to planetary-mass companions cannot reach the centre where they can break\footnote{For stellar-mass companions it is possible for IGWs to break above the convective core \citep[See Fig.~9 in][]{barker_tidal_2020}.} and are instead reflected from the convective core interface, typically leading to inefficient dissipation \citep[e.g.][]{barker_three-dimensional_2011,barker_tidal_2020}.

Of particular interest is the observed orbital decay of WASP-12 b \citep{maciejewski_departure_2016,patra_apparently_2017,maciejewski_planet-star_2018,yee_orbit_2019} which suggests $Q^\prime \approx 2\times 10^5$ for the stellar host \citep{patra_continuing_2020}.
If WASP-12 is a subgiant star, then existing theories of nonlinear wave breaking near the core can produce efficient tidal dissipation ($Q^\prime\approx 2.7\times 10^5$, \citealt{weinberg_tidal_2017,bailey_understanding_2019,barker_tidal_2020}). 
However, observed stellar properties do not favour such subgiant models \citep{hebb_wasp-12b_2009,bailey_understanding_2019,akinsanmi_tidal_2024,leonardi_taste_2024}. 
If WASP-12 is a main-sequence star, then it would have a convective core, so existing theories predict weak tidal dissipation. Intriguingly, \citet{barker_tidal_2020} showed that if there was some other mechanism to dissipate the tidally-excited IGWs, main-sequence models could have efficient tidal dissipation with $Q^\prime\approx 2.2\times 10^5$ (depending on mass), roughly consistent with observations.
Can we explain the observed $Q^\prime$ if the star has a convective core?

Many other proposed tidal dissipation mechanisms in stars with convective cores have been shown to be inefficient. These include:
convective turbulence acting on the large-scale equilibrium tide \citep[e.g.][]{zahn_les_1966,goldreich_turbulent_1977,
ogilvie_interaction_2012,duguid_tidal_2019,duguid_convective_2020,vidal_efficiency_2020,vidal_turbulent_2020}; the elliptical instability \citep[e.g.][]{devries_tidal_2023}; (linearly-excited) inertial waves, which are not excited unless the tidal period is longer than half the stellar rotation period \citep[e.g.][]{ogilvie_tidal_2007,ivanov_inertial_2010,ogilvie_tides_2013,mathis_variation_2015,bolmont_effect_2016,barker_tidal_2020,AB2023}; or resonance locking that typically operates on the stellar evolution timescale, ${\sim}\text{Gyr}$ \citep[e.g.][]{ma_orbital_2021}.
While recent work has explored alternative mechanisms for critical layer formation which would justify the fully damped IGW regime \citep[e.g.~by linear radiative damping,][]{Guo2023}, the impact of magnetic fields on tidally excited IGWs remains unexplored to date.

\begin{figure}[htb!]
	\plotone{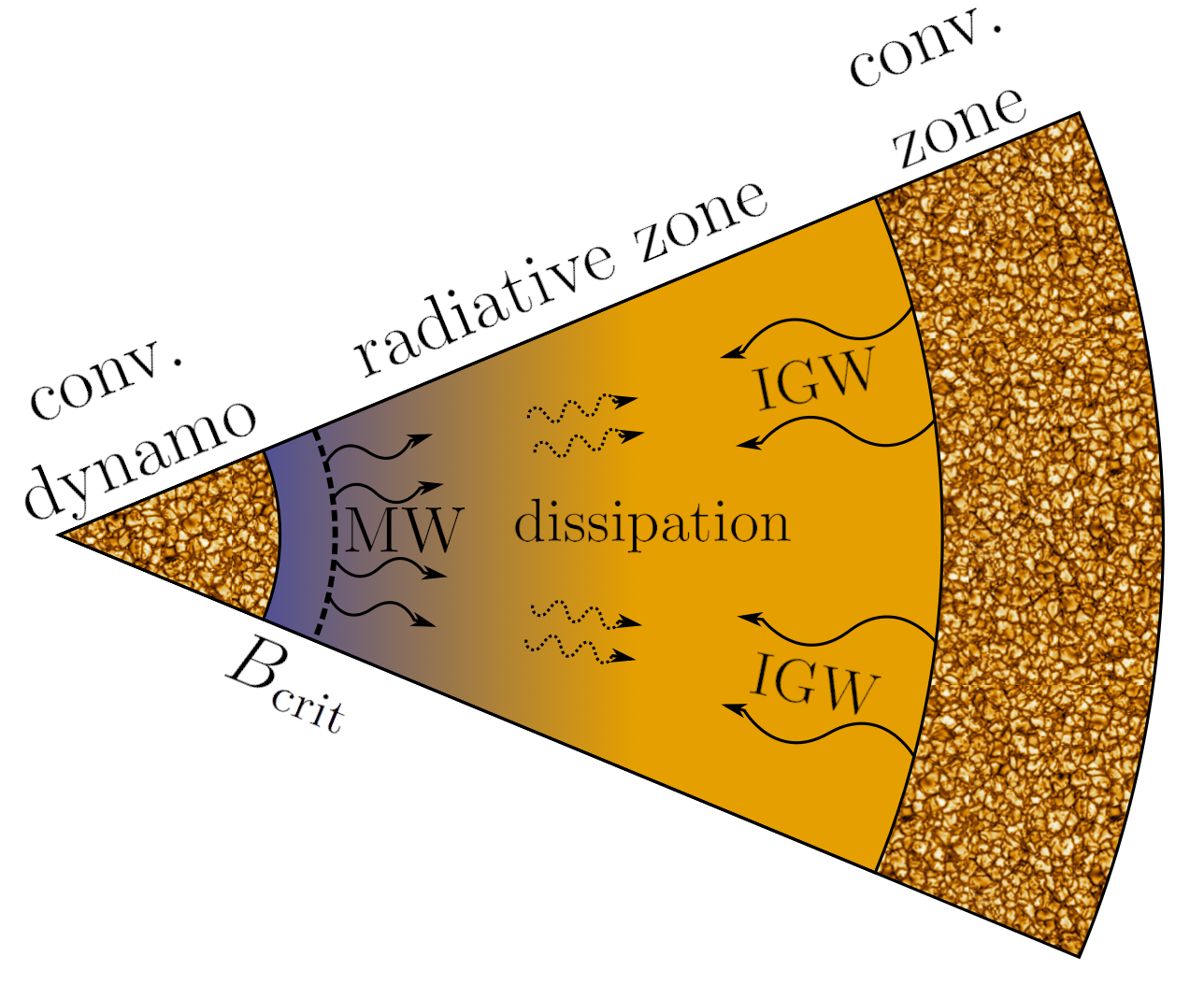}
	\caption{Schematic of tidal dissipation due to wave conversion. Tidally excited internal gravity waves (IGWs) are launched from the boundary between the convective envelope and radiative zone and travel towards the 
  centre of the star. If a convective core dynamo produces a field strong enough to surpass a critical threshold $B_\text{crit}$, then the IGWs are fully converted to outwardly propagating magnetic waves (MWs). The outward MWs can then be fully dissipated as they propagate through the magnetized radiative zone.}
\label{fig_schematic}
\end{figure}

Motivated by astroseismic observations of red giant stars \citep{Stello2016},
\citet{fuller_asteroseismology_2015} suggested magnetic fields can strongly impact the propagation of IGWs in stars.
Subsequent work \citep{lecoanet_conversion_2017,lecoanet_asteroseismic_2022,Rui2023} has shown that if the IGWs encounter a  sufficiently strong vertical (radial) magnetic field they are fully converted into outwardly propagating magnetic waves (MWs), which are subsequently fully dissipated.
Though note that others find weaker interactions \citep{Loi2017,Loi2018,Loi2020b}, which only partially damp the waves 
\citep{Mosser2017}.
In the tidal context (see Fig.~\ref{fig_schematic} for a schematic representation), tidally excited IGWs from the convective envelope boundary propagate through the radiative zone towards a convective core.
Since a convective core is thought to be conducive to dynamo action \citep{charbonneau_magnetic_2001,browning_simulations_2004,brun_simulations_2005}, it could be expected that field will leak into the lower regions of the radiative zone (e.g.~when the convective core retreats or via ohmic/turbulent diffusion).
If the field strength is sufficient, then wave conversion can operate and IGWs are fully converted into outwardly travelling MWs. As these MWs propagate through the radiative zone into regions of weaker field their local radial wavenumbers increase rapidly until they are fully damped by radiative or ohmic diffusion.

While wave conversion could result in efficient tidal dissipation it is unclear if the criterion for wave conversion is satisfied in observed systems. 
In particular, this mechanism has the potential to reconcile the original and recent observations
of \cite{hebb_wasp-12b_2009,akinsanmi_tidal_2024} (inferring WASP-12 to be a main-sequence star) with observational constraints on $Q^\prime$ for WASP-12 b. 
Motivated by this problem, in this Letter we use stellar evolutionary models to evaluate the critical field strength required for tidally excited IGWs to be converted into MWs.
In \S~\ref{sec_wave_conversion} we describe the wave conversion criterion and the models used in this study, before computing $Q^\prime$ and the resulting inspiral timescales for a prototypical hot Jupiter planet in \S~\ref{sec_Q_and_tau}. Finally, in \S~\ref{sec_discussion_conclusions} we briefly discuss our results and conclude.

\section{Wave conversion} \label{sec_wave_conversion}

In order to assess the viability of wave conversion to provide efficient tidal dissipation we construct a set of  stellar models using the stellar evolution code MESA \citep[version r22.11.1,][]{Paxton2011, Paxton2013, Paxton2015, Paxton2018, Paxton2019}. Our models cover a range of  main sequence stars with masses $M_\star \in \{1.2, 1.3, 1.4, 1.5, 1.6\}M_\odot$. Stars in this mass range possess a convective envelope, where tidally excited IGWs can be launched from, as well as a convective core for significant portions of their main sequence lifetime, where we can anticipate a convective dynamo to be (or have been) operating.
Since we are interested in evaluating the wave conversion mechanism over the lifetime of the star we adopt the Schwarzschild criterion for convective instability \citep{anders_schwarzschild_2022}. While mixing at the base of the radiative zone may be important, mixing processes remain an area of great uncertainty in stellar models (see e.g.~\citealt{kupka_modelling_2017,aerts_probing_2021} and references therein).
In the stellar models presented here, we neglect additional sources of mixing such as convective overshooting. We also analyzed similar models run with standard convective overshoot schemes, and found similar results.
The mixing length parameter is not important for core convection \citep{joyce_review_2023} so we adopt the typical value of $\alpha_\text{MLT} = 2$.
The stellar metallicity is chosen to be $Z=0.02$.

The criterion for wave conversion is that the local radial magnetic field strength $\lvert B_r \rvert$ is greater than some critical threshold at which the radial wavenumber of IGWs and MWs match \citep{fuller_asteroseismology_2015,lecoanet_conversion_2017,lecoanet_asteroseismic_2022, Rui2023},
\begin{equation}
 \lvert B_r \rvert > B_{\mathrm{crit}} \,, \label{eq_wave_criteria}
\end{equation}
with $B_{\mathrm{crit}}$ defined within the radiative zone as
\begin{equation} \label{maths_Bcrit}
	B_{\mathrm{crit}}(r,t) =  \frac{ \pi \omega^2 r \sqrt{\rho}}{N} 
\end{equation}
where $\omega$ is the wave frequency (tidal frequency in our context), $r$ is the radius from the centre of the star, $\rho$ is the density, and $N$ is the Brunt--V\"ais\"al\"a frequency (note that $\sqrt{N^2}$ takes real values only within radiative zones).

\begin{figure}[htb!]
	\plotone{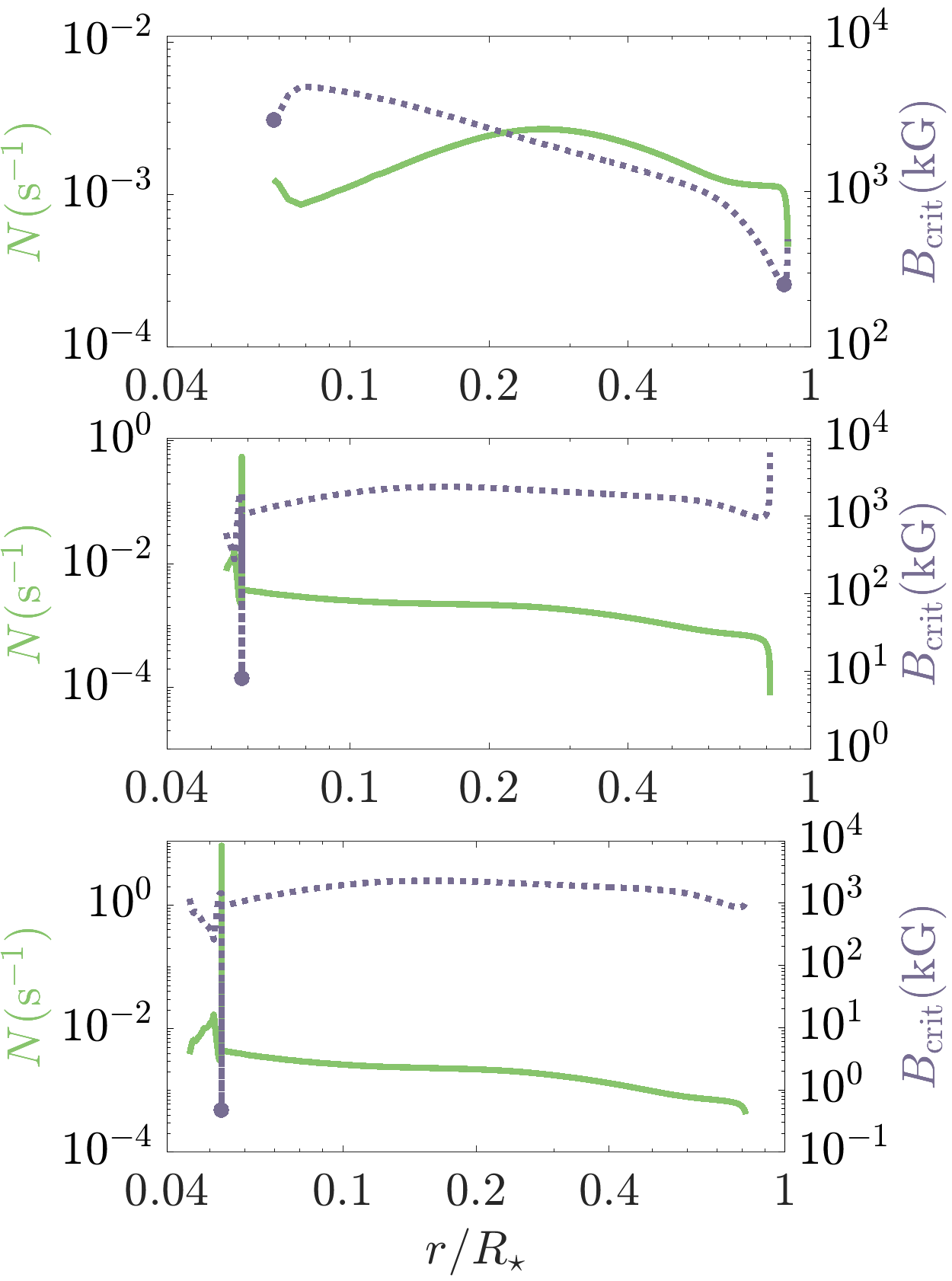}
	\caption{Radial profiles within the radiative zone of the buoyancy frequency, $N(r)$, and the critical field strength for wave conversion, $B_{\mathrm{crit}}(r)$, for our $1.3M_\odot$ stellar model. We show these quantities for three different times $t \in \{0.23, 2.14, 3.3 \}\times10^9$ years (corresponding to core hydrogen mass fractions $X \approx\{0.66, 0.31, 0.02\}$). These correspond to prior to convective core retreat, some time after the convective core begins to shrink leading to a localised spike in $N$, and some time after the convective core has vanished. The latter of these highlights the persistence 
    of the $N$-spike. 
    Solid dots on the $B_{\mathrm{crit}}$ lines denote the global and local minima within the lowest 10\% of the radiative zone. We use the latter 
    for our evaluation of $B_{\mathrm{crit}}$ throughout this work. 
}
	\label{fig_N2}
\end{figure}

In Fig.~\ref{fig_N2} we show radial profiles of $B_{\mathrm{crit}}(r,t)$ within a $1.3 M_\odot$ star at fixed times $t \in \{0.23, 2.14, 3.3 \}$ Gyrs for a 0.5 day tidal period (${\approx}1$ day orbital period for a tide raising body orbiting a slowly rotating star).
At early times we see the global minimum located near the convective envelope boundary.
As the star evolves, the convective core retreats leaving behind products of nuclear burning and thereby building up the composition gradient and locally steepening the stratification, which we see as an increase in $N$.
Since $N$ is a key parameter in the denominator of Eq.~\ref{maths_Bcrit}, this local increase in $N$ results in a significant reduction of $B_{\mathrm{crit}}$ just above the convective core.
The coincidence of the local maximum in $N$ and minimum of $B_{\mathrm{crit}}$ can be seen in Fig.~\ref{fig_N2}, which includes example profiles of $N$ from our models.

If wave conversion occurs at all, it can be expected to take place at the largest $r$ within the radiative zone such that Eq.~\ref{eq_wave_criteria} is satisfied. 
However, while wave conversion is most susceptible when $B_{\mathrm{crit}}(r)$ takes its minimum, it is clear from Fig.~\ref{fig_N2} that the global minimum may be located far from a convective core dynamo.
As such, we limit our region of evaluation of $B_{\mathrm{crit}}$ to where we believe wave conversion is most likely to occur. We define this as the lower 10\% of the radiative zone (which in practice may extend to the centre if no convective core exists) and hence close to an assumed convective core dynamo.

\begin{figure*}[htb!]
	\plotone{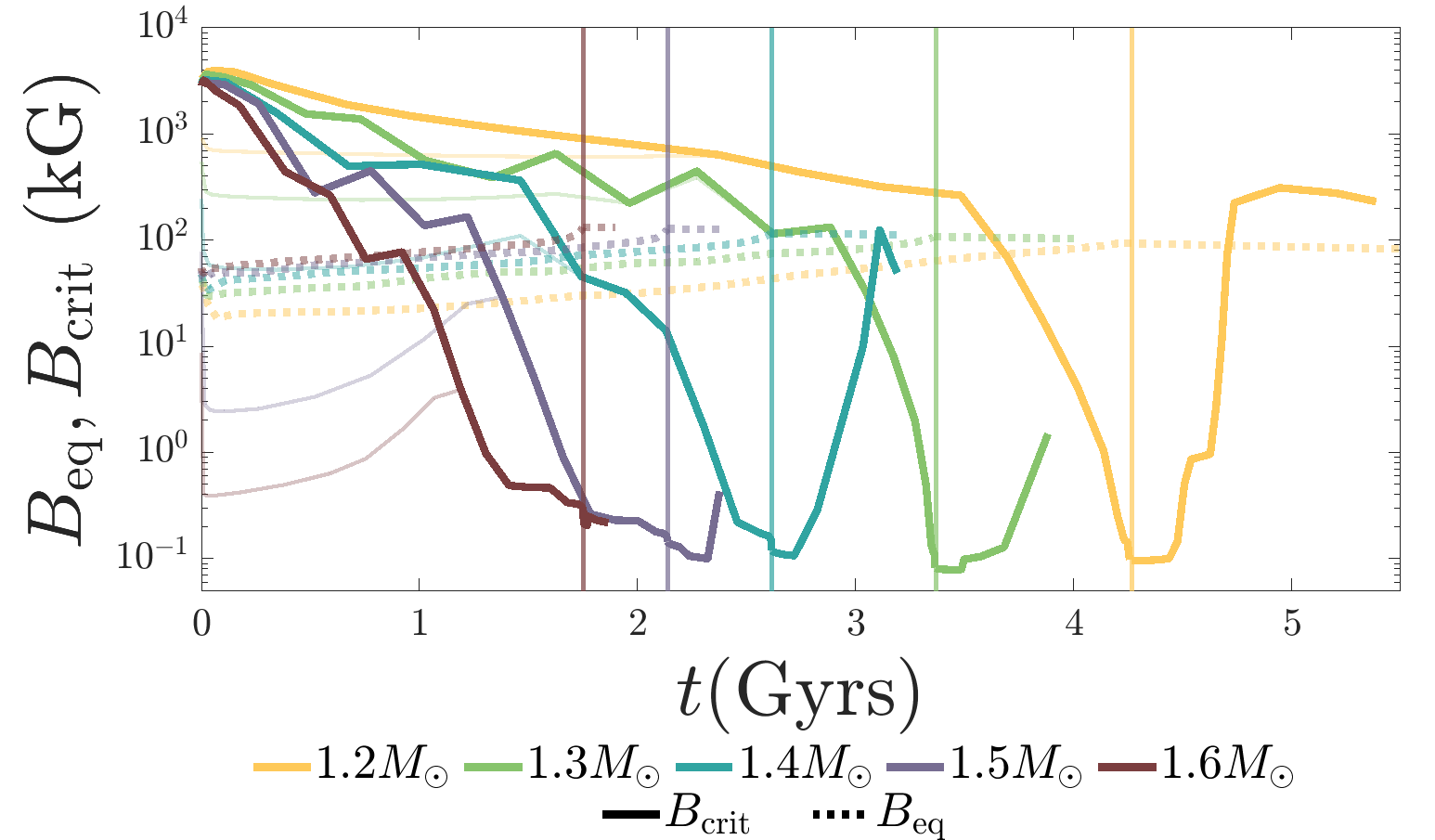}
	\caption{Time series for each stellar model $M_\star \in \{1.2, 1.3, 1.4, 1.5, 1.6\}M_\odot$ (see legend) of: the estimated convective core dynamo field strength $B_\text{eq}$ using equipartition arguments, and the threshold for wave conversion $B_\text{crit}$ as computed by Eq.~\ref{maths_Bcrit}. The thinner branch of $B_\text{crit}$ denotes the value if Eq.~\ref{maths_Bcrit} is evaluated using the global minimum of $B_\text{crit}$ rather than the local minimum in the lower 10\% of the radiative zone. The vertical lines indicate when the convective core vanishes for each respective model. Note that when $B_\text{eq} > B_\text{crit}$ then we can expect conversion of IGWs to MWs and hence efficient tidal dissipation.}
	\label{fig_bcrit_b}
\end{figure*}

For the left hand side of Eq.~\ref{eq_wave_criteria}, we estimate the radial field strength of the convective core $\lvert B_r \rvert$ through the assumption of equipartition of the magnetic and convective kinetic energy densities ($\lvert \boldsymbol{B} \rvert {\sim} \lvert B_r \rvert {\sim} B_\text{eq}$). We define the equipartition magnetic field strength as the volume-averaged (mean) kinetic energy within the convective core,
\begin{equation}\label{maths_B_eq}
	B_\text{eq}(t) = \frac{6\sqrt{\pi}}{R_{\text{c.c.}}^3} \int_{0}^{R_{\text{c.c.}}} r^2 \sqrt{\rho u_{\text{conv}}^2}\, \text{d} r  \,,
\end{equation}
where $u_\mathrm{conv}$ is the convective velocity and $R_{\text{c.c.}}$ is the radius of the convective core. 
As can be seen in Fig.~\ref{fig_bcrit_b} (dotted lines), using this estimate we find that all our models result in field strengths on the order of 10-100 kG 
which we note is in good agreement with numerical simulations of convective core dynamos (e.g.~\citealt{brun_simulations_2005}, albeit for A-Class $2 M_\odot$ stars).
While there is substantial uncertainty in the typical lengthscale of these dynamo fields, \citet{lecoanet_conversion_2017} found that the critical magnetic field strength for wave conversion (Eq.~\ref{maths_Bcrit}) only changes by a factor of two when considering large-scale (field varying on the scale of the wave) vs.~small scale (field variation length a factor of 50 smaller than the wavelength) magnetic fields.
Furthermore, it is possible for the dynamo-generated field to be in a state of super-equipartition, depending on the saturation mechanism. 
In particular, a ten-fold increase in the equipartition estimate can be obtained by the presence of a fossil field \citep{featherstone_effects_2009} due to a magnetostrophic balance. 
Further, such fossil fields themselves can possess field strengths in the range of 10-100kG \citep{brun_magnetism_2017}.
As such, our estimate using the equipartition field strength could plausibly be treated as a lower limit.

After the convective core vanishes, nonlinear wave breaking could also cause efficient tidal dissipation. However, a remnant magnetic field from the convective dynamo will still be present in the core of the star \citep[e.g.,][]{fuller_asteroseismology_2015}. Depending on the strength of the remnant magnetic field, the mechanism of tidal dissipation may be nonlinear wave breaking, or conversion to MWs. In both cases, $Q^\prime$ will be similar because the waves will be approximately fully dissipated.
We assume the magnetic fields diffuse ohmically \citep{spitzer_transport_1953,bellan_fundamentals_2008} following the end of core convection to extend the field strength line in Fig.~\ref{fig_bcrit_b} beyond the time the convective core vanishes.

Also shown in Fig.~\ref{fig_bcrit_b} are time series of the critical field strength for wave conversion $B_{\mathrm{crit}}(t)$ for each of our stellar models.
Note that we include evaluation of $B_\text{crit}$ using the global minimum value of $B_\text{crit}$ for comparison with the local minimum, although the location of $B_\text{crit}$  using this method is typically towards the upper portions of the radiative zone and hence far from a convective core dynamo during the early stages of stellar evolution.
In all our models, while the star has a convective core $B_{\mathrm{crit}}$ decays over approximately four orders of magnitude, with a faster rate for the more massive stars whose structures evolve more quickly.
The rate is roughly constant for $M_\star \in \{1.4, 1.5, 1.6\}M_\odot$ while for $M_\star \in \{1.2, 1.3\}M_\odot$ the rate is roughly constant until approximately 0.5 Gyr before the convective core vanishes, at which point there is a rapid reduction of $B_{\mathrm{crit}}$. Once the convective core has vanished $B_\text{crit}$ increases rapidly until the termination of our models.
For the $1.2M_\odot$ model $B_\text{crit}$ then saturates at the end of the main sequence.

From Fig.~\ref{fig_bcrit_b} it can be seen that while Eq.~\ref{eq_wave_criteria}, using the local minimum of $B_\text{crit}$, is not initially satisfied in any model, at times $t \approx \{3.9, 2.7, 1.7, 1.3, 0.7 \}$ Gyr for $M_\star \in \{1.2, 1.3, 1.4, 1.5, 1.6\}M_\odot$, respectively, $B_\text{crit}$ becomes larger than $B_\text{eq}$. 
This indicates that wave conversion operates beyond these times for a given mass\footnote{It should be noted that if a sufficient argument could be made for stars with masses $M_\star \gtrsim 1.5 M_\odot$ to possess strong fields in the upper radiative zone then it is possible for such stars to satisfy Eq.~\ref{eq_wave_criteria} throughout their entire lifetimes.}.

Due to the slow ohmic decay of the field, wave conversion is able to continue even once the convective core has vanished. For stars with $M_\star > 1.2M_\odot$ wave conversion continues for the remaining duration of the main sequence while for the $1.2M_\odot$ case $B_\text{crit}$ raises above $B_\text{eq}$, suggesting the mechanism shuts off. These results suggest that wave conversion can (and is more likely to) operate in the latter stages of the main sequence lifetime and that it can persist for Gyr timescales.

If the field is in a state of super-equipartition then the field strength could be more than an order of magnitude larger
than our equipartition estimate. In which case the duration that wave conversion can occur could be significantly longer than Fig.~\ref{fig_bcrit_b} suggests.
Further, for a super-equipartition field in the $1.2 M_\odot$ case, since $B_\text{crit}$ plateaus to a finite amplitude, wave conversion could continue to operate until the end of the main sequence.
Besides the convectively-generated field strength, there are further uncertainties in the radial profile of $N$. 
While the exact profile of $N(r)$ is unknown, large peaks are generic in stellar evolution models (with and without convective overshoot), and can be inferred from asteroseismic observations of more massive stars \citep[e.g.][]{pedersen_internal_2021}.
Nevertheless, these uncertainties only act to shift the time at which we can expect wave conversion to operate to earlier (for larger $B_\text{eq}$ or $N$ values) or later times (for smaller $B_\text{eq}$ or $N$).

Fig.~\ref{fig_bcrit_b} has demonstrated that wave conversion can continue to operate even after the convective core has vanished. However, with the convective core gone, it is now more likely that wave breaking of the IGWs can occur in the radiative core that remains. For both mechanisms inwardly-propagating IGWs do not proceed beyond the largest radius where they operate, since wave conversion fully converts waves to outwardly propagating MWs at this location, and wave breaking typically leads to complete absorption of IGWs. Hence, if both mechanisms can operate then the one that occurs at the largest $r$ is the one that will take precedence. 

The expected radius for which IGWs will non-linearly break can be estimated by computing $\lvert \xi_r k_r \rvert$ from WKB theory \citep{goodman_dynamical_1998,weinberg_tidal_2017,barker_tidal_2020}, where $\xi_r$ is the radial displacement of the IGW (more specifically its $l=m=2$ spherical harmonic coefficient, averaged over a wave period) and $k_r$ is the radial wavenumber of the IGWs. When $\lvert \xi_r k_r \rvert \gtrsim 1$ in the radiative zone then wave breaking can be expected. Strong non-linear interactions are expected for even lower amplitudes but are more uncertain to estimate \citep[e.g.][]{Weinberg2024}. 
Wave conversion occurs at the largest $r$ such that Eq.~\ref{eq_wave_criteria} is satisfied. A subtlety in our evaluation of $B_\text{crit}$ is that we are seeking where wave conversion is most likely to take place, not a strict evaluation of the largest $r$ such that Eq.~\ref{eq_wave_criteria} is satisfied. 
Essentially, we are calculating the smallest $r$ such that if the inwardly propagating IGWs have not been converted by this point they will certainly not be converted below this $r$. This plausibly provides a sufficiently strict test of which mechanism takes precedence.
We find that in all cases for which the strict criterion for wave breaking $\lvert \xi_r k_r \rvert > 1$ is satisfied, this occurs at fractional radii $r/R_\star$ which are approximately two orders of magnitude smaller than where we can expect wave conversion.
Hence, we anticipate that efficient tidal dissipation due to wave conversion will continue to operate preferentially over nonlinear wave breaking at some finite, non-zero $r$ for a prolonged period of time, even after the convective core has vanished.

\section{The tidal quality factor and inspiral timescale} \label{sec_Q_and_tau}

\begin{figure*}[htb!]
	\plotone{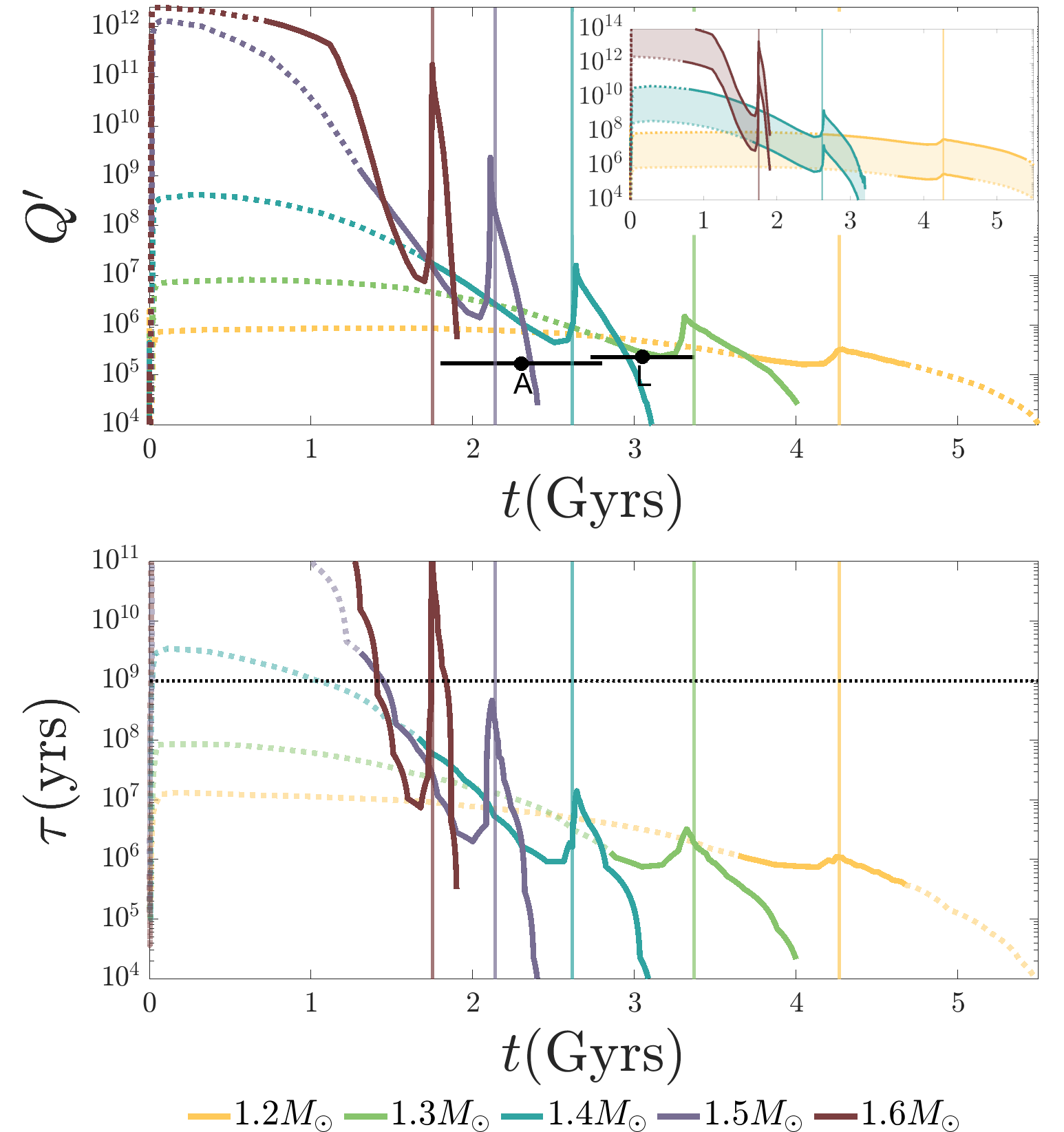}
	\caption{Time series for each stellar model $M_\star \in \{1.2, 1.3, 1.4, 1.5, 1.6\}M_\odot$ (see legend) of the dissipation due to wave conversion parametrised by the modified tidal quality factor $Q^\prime$ ($1/Q^\prime$ is proportional to the dissipation). Each $Q^\prime$ curve is partitioned into a region where we expect wave conversion to operate (solid) and where the criterion is not satisfied (dotted). Note that this partitioning assumes the core dynamo-generated field is in equipartition. 
	If the field strength is above equipartition then the period where wave conversion can operate is extended.
	The vertical lines denote when the convective core vanishes for each respective model. The insert in the top panel displays the same information but the upper curve for each respective model is for a 1.5 day tidal period (3 day orbit). 
	Included in the top panel are observational estimates of $Q^\prime$ taken from \cite{akinsanmi_tidal_2024} and \cite{leonardi_taste_2024} (point A and L respectively) for WASP-12b where the assumed stellar masses are $\{1.422, 1.325\}M_\odot$ respectively. Note that error bars for $Q^\prime$ are within the symbol.
	The lower panel shows the inspiral timescale associated with each stellar model for a Jupiter mass planet on a one day orbit.
	The same labelling as for the upper panel is adopted. 
}
	\label{fig_Q}
\end{figure*}

Once the IGWs are converted to MWs they are likely to be fully damped as their local radial wavenumber will become extremely large as they propagate outward into regions of weaker magnetic fields. 
For fully damped IGWs, we can simply estimate the resulting (modified) tidal quality factor\footnote{We ignore the effects of magnetic fields on the wave excitation from the radiative/convective envelope boundary. This is reasonable as the field strengths required to substantially modify wave launching are presumably comparable with $B_\text{crit}$, and hence are unlikely to be achieved there in these stars.} $Q^\prime$ quantifying the dissipation using our stellar models (e.g.~\citealt{zahn_dynamical_1975,goodman_dynamical_1998,ogilvie_tidal_2007,barker_three-dimensional_2011,chernov_dynamical_2017,barker_tidal_2020,Ahuir2021}). 
The precise damping mechanism of the MWs is unimportant as long as the waves are fully damped so that they can be modelled as ``travelling waves" propagating inwards from the radiative/convective boundary.
In Fig.~\ref{fig_Q} we show how $Q^\prime(t)$ varies for each of our stellar models \citep[using Eq.~41 of][]{barker_tidal_2020}, assuming the IGWs are  tidally excited by a Jupiter-mass planet on a one-day orbit around a slowly rotating star (giving a tidal period $P_\mathrm{tide}\approx P_\mathrm{orbit}/2=0.5$ d). Highlighted in Fig.~\ref{fig_Q} is when we expect tidal dissipation due to wave conversion to operate.

The temporal profiles of $Q^\prime$ for each stellar mass are qualitatively similar with a larger value (indicating weaker dissipation) at early times that reduces to a local minimum (indicating maximum dissipation) just before the core vanishes, at which point there is a brief spike before $Q^\prime$ reduces further.
For the more massive stars this trend is more pronounced. For the $1.6 M_\odot$ model there is a 5 orders of magnitude drop from $Q^\prime{\sim}10^{12}$ at early times to $Q^\prime{\sim}10^{7}$ just before the convective core vanishes.
In contrast, the $1.2 M_\odot$ case remains within the range $Q^\prime\in [10^{5}, 10^6]$ throughout its lifetime.

While these estimates have assumed a fixed 1 day orbital period (0.5 day tidal period), the wave conversion mechanism can occur for a range of orbital periods.
Wave conversion occurs when the magnetic field surpasses $B_\text{crit} \propto \omega^2 \propto P_\text{orbit}^{-2}$ (for slow rotation).  It is clear then that wider orbits result in a smaller $B_\text{crit}$ and hence the stellar interior is more prone to wave conversion. 
Indeed, if the orbital period is increased to 3 days (1.5 day tidal period) we find that wave conversion can be expected (under the assumption of an equipartition field) for all our models over significantly larger portions of their main sequence lifetimes. However, any change to the orbital period also has an impact on the tidal quality factor, which is strongly dependent on it, scaling as $Q^\prime \propto P_\text{orbit}^{8/3}$.
As such, while increasing the orbital period widens the window of time within which wave conversion can operate it leads to 
less efficient tidal dissipation (larger $Q^\prime$).
This is demonstrated within the insert of Fig.~\ref{fig_Q} where we can see for the $1.2M_\odot$ star that a 3 day orbital period results in wave conversion operating  for around three times as long as the 1 day case, but the value of $Q^\prime$ is approximately two orders of magnitude larger.

While $Q^\prime$ is a useful parameter to determine the efficiency of tidal dissipation, the details of the star itself, and the planetary mass and orbital period, are important to determine the resulting inspiral timescale for orbital decay of a close-in planet. 
An inspiral 
timescale $\tau$ can be estimated, under the assumption of constant (in time) $Q^\prime$, by integrating the tidal evolution equation for the semi-major axis of the secondary (see e.g.~\citealt{murray_solar_2000,ogilvie_tidal_2014,barker_tidal_2020}, in particular Eq.~56-57 of the latter). 
We show in the bottom panel of Fig~\ref{fig_Q} the resulting $\tau$ from our models for a Jupiter mass planet on a one day orbit around its host star, highlighting when we can expect
tidal dissipation due to wave conversion to occur. 
For stars with $M_\star < 1.6M_\odot$, when wave conversion is in operation, the inspiral timescale is shorter than a Gyr for a 1 day orbit. The $1.6M_\odot$ case 
features a brief spike above the Gyr timescale immediately after the convective core vanishes, but $\tau$ is generally below 1 Gyr while this mechanism operates. As such, not only do all stars in our mass range feature periods of time where tidal dissipation due to wave conversion can occur, the resulting timescales for the inspiral of a closely-orbiting giant planet are sufficiently small to be astrophysically relevant.

Since $B_\text{crit} \propto P_\text{orbit}^{-2}$, if the planet's orbit decays rapidly enough it is possible that $B_\text{crit}$ becomes large enough to shut off the wave conversion mechanism. As such, the rate of orbital decay due to wave conversion may be constrained by the rate of change of $B_\text{crit}$ as the planet's orbit shrinks. However, in the absence of a more detailed orbital evolution model it is unclear if such a constraint is relevant.

\section{Discussion and conclusions} \label{sec_discussion_conclusions}

Using stellar models we have demonstrated that the wave conversion mechanism, where inwardly-propagating tidally-excited IGWs are converted to outwardly-propagating MWs, can operate for significant fractions (on the order of Gyrs) of the main sequence lifetimes of stars with masses in the range $[1.2- 1.6]M_\odot$.
Since the MWs are fully damped as they travel through the radiative zone, wave conversion constitutes a new mechanism to enter the fully-damped regime for tidally-excited IGWs. 
Importantly, since wave conversion can occur both while and after a convective core exists, this extends the duration of the fully-damped IGW regime beyond what is predicted by wave breaking alone. Furthermore, this duration could be significantly extended beyond what the presented models suggest if the field produced by a convective core dynamo is stronger than the one expected from equipartition of convective kinetic and magnetic energy densities.

In this work, we have assumed full conversion from IGWs into MWs which are subsequently damped; thus, this mechanism of IGW dissipation gives the same tidal dissipation rates as wave breaking where the waves are also fully damped.
If the IGWs are only partially converted into MWs, the tidal dissipation rate would plausibly be smaller than our estimates by the conversion rate.
As works which do not find perfect conversion still find conversion rates of ${\sim} 80\%$ \citep[e.g.,][]{Loi2018}, this uncertainty does not significantly alter our analysis.

While the tidal dissipation rate for this mechanism is similar to that of wave breaking, 
the energy is deposited in a very different location.
For non-linear wave breaking of IGWs the energy is deposited typically at small $r$ in the form of kinetic energy (though there is mixing of entropy and some viscous heating) and there is a transfer of angular momentum spinning up the star from the inside out as the planet's orbit decays. In contrast, the converted MWs propagate outwards and are dissipated by radiative or ohmic diffusion in the radiative zone. 
As a result, the angular momentum deposited inside the star by the damping of these waves may occur at a much larger radius.

Before turning our attention to the application of this mechanism to hot Jupiters, it is worth highlighting that the underlying theory behind wave conversion is linear. Hence, unlike for wave breaking of IGWs, there is no minimum mass threshold imposed on the tide raising body for this mechanism to operate. As such, tidal dissipation due to wave conversion is applicable to all planetary masses, and in particular it applies to ultra short period planets (USPs) \citep{Hamer2020}. A crude estimate for the inspiral timescale based on Fig.~\ref{fig_Q} can be evaluated by multiplying $\tau$ by $1/M_\text{p}$ where $M_\text{p}$ is the mass of the planet in Jupiter masses. This suggests that even Earth-mass planets could have inspiral timescales shorter than 1 Gyr during the dips in $\tau$ as the star evolves off the main sequence. This mechanism may therefore play a role in the tidal orbital decay of USPs.

Returning to the issue of WASP-12b, the main-sequence models (containing convective cores) in \cite{bailey_understanding_2019} were found to be a closer match to observational data.
In \cite{barker_tidal_2020} the main-sequence model that most closely agreed with the observed value of $Q^\prime$ was for a $1.32M_\odot$ model for WASP-12, which possesses a convective core. 
Since these models, observations have been able to further constrain the stellar parameters of WASP-12 \citep{akinsanmi_tidal_2024,leonardi_taste_2024}, which provide age ($\{2.3^{+0.5}_{-0.5}, 3.05^{+0.32}_{-0.32}\}$ Gyr respectively), mass ($\{1.422^{+0.077}_{-0.069},1.325^{+0.026}_{-0.018}\} M_\odot$ respectively) and tidal quality factor ($Q^\prime \approx \{1.7^{+0.14}_{-0.14} ,2.13^{+0.18}_{-0.18}\}\times 10^5$ respectively) estimates.
Figs.~\ref{fig_bcrit_b} and \ref{fig_Q} suggest that wave conversion can occur within the estimated age window and mass range of these observations even in presence of a convective core.
Our results find excellent agreement with the observationally constrained value of $Q^\prime$ of \cite{leonardi_taste_2024} although it should be cautioned that the metallicity of our models differs from that used in \cite{leonardi_taste_2024} ($Z\approx0.03$).
This mechanism is therefore a promising one for enabling efficient dissipation of tidally excited gravity waves in stars with convective cores, which might resolve the existing observational conundrum for WASP-12.

There have been efforts to predict $Q^\prime$ from tidal theory \citep{barker_tidal_2020,ma_orbital_2021} and to constrain its value from transit timing variations (TTV) in hot Jupiter systems \citep[e.g.][]{patra_continuing_2020,Maciej2020,Maciej2022,Basturk2022,Shan2023}. \cite{barker_tidal_2020} and \cite{ma_orbital_2021} estimated $Q^\prime$ in the fully-damped IGW regime for a number of exoplanetary systems. In particular WASP-114b, WASP-122b, KELT-16b, and OGLE-TR-56b were all noted to have convective cores but small enough $Q^\prime$ that orbital decay could potentially be observed through TTVs using observations spanning about a decade\footnote{We also highlight TOI-2109b as another promising candidate for detection of tidally-driven TTVs.}. For these cases the estimated stellar ages are such that wave conversion could indeed be taking place and hence further observations of these systems would help constrain tidal theory as well as the validity of the mechanism we have proposed here.
These applications illustrate the potential importance of this new magnetically-mediated tidal dissipation mechanism.

\begin{acknowledgments}
CDD was supported by STFC grant ST/X001083/1 and
a Research Project Grant from the Leverhulme Trust RPG-2020-109.
NBV was supported by EPSRC studentship 2528559. 
DL is supported in part by NASA HTMS grant 80NSSC20K1280 and NASA OSTFL grant 80NSSC22K1738. 
AJB was supported by STFC grants ST/S000275/1 and ST/W000873/1. 
AJB would like to thank the Isaac Newton Institute for Mathematical Sciences, Cambridge, for support and hospitality during the programme ``Anti-diffusive dynamics: from sub-cellular to astrophysical scales" where part of the work on this paper was undertaken. This work was supported by EPSRC Grant No.~EP/R014604/1. 
We are grateful for the time and effort of the anonymous referee and their useful comments that helped improve this paper.

The data is available on Zenodo under an open-source Creative Commons Attribution license: 
\dataset[doi:10.5281/zenodo.10858577]{https://doi.org/10.5281/zenodo.10858577}.
\end{acknowledgments}

\software{MESA \citep{Paxton2011, Paxton2013, Paxton2015, Paxton2018, Paxton2019}
          }

\bibliography{IGW_alfven}{}
\bibliographystyle{aasjournal}

\end{document}